\documentclass[preprint2]{aastex}

\shorttitle{Testing MOND in Pal~14}

\begin{document}

 \title{Testing Fundamental Physics with Distant Star Clusters:\\
Analysis of Observational Data on Palomar~14\altaffilmark{*,**,***}}

\author{K. Jordi\altaffilmark{1,2}, E.~K. Grebel\altaffilmark{2}, M. Hilker\altaffilmark{3}, H. Baumgardt\altaffilmark{4}, M. Frank\altaffilmark{2}, P. Kroupa\altaffilmark{4}, H. Haghi\altaffilmark{4,5}, P. C{\^o}t{\'e}\altaffilmark{6}, and S.~G. Djorgovski\altaffilmark{7}}

\altaffiltext{*}{Based on observations made with ESO Telescopes at the Paranal Observatories under programme ID 077.B-0769}
\altaffiltext{**}{Based on observations made with the NASA/ESA Hubble Space Telescope, obtained from the data archive at the Space Telescope Institute. STScI is operated by the association of Universities for Research in Astronomy, Inc. under the NASA contract  NAS 5-26555.}
\altaffiltext{***}{Some of the data presented herein were obtained at the W.M. Keck Observatory, which is operated as a scientific partnership among the California Institute of Technology, the University of California and the National Aeronautics and Space Administration. The Observatory was made possible by the generous financial support of the W.M.~Keck Foundation.}
\altaffiltext{1}{Astronomisches Institut, Universit\"at Basel, Venusstrasse 7, CH-4102 Binningen, Switzerland}
\altaffiltext{2}{Astronomisches Rechen-Institut, Zentrum f\"ur Astronomie der Universit\"at Heidelberg, M\"onchhofstrasse 12~-~14, D-69120 Heidelberg, Germany}
\altaffiltext{3}{European Southern Observatory, Garching b. M\"unchen, Germany}
\altaffiltext{4}{Argelander Institut f\"ur Astronomie, Auf dem H\"ugel 71, D-53121 Bonn, Germany}
\altaffiltext{5}{Institute for Advanced Studies in Basic Sciences (IASBS), P.O.Box 45195-1159, Zanjan, Iran}
\altaffiltext{6}{Herzberg Institute of Astrophysics, National Research Council of Canada, Victoria, BC V9E 2E7, Canada}
\altaffiltext{7}{Astronomy Department, California Institute of Technology, Pasadena, CA~91125, USA}

\begin{abstract}
We use the distant outer halo globular cluster Palomar~14 as a test case for classical vs. modified Newtonian dynamics (MOND). Previous theoretical calculations have shown that the line-of-sight velocity dispersion predicted by these theories can differ by up to a factor of three for such sparse, remote clusters like Pal~14. We determine the line-of-sight velocity dispersion of Palomar~14 by measuring radial velocities of 17 red giant cluster members obtained using the Very Large Telescope (VLT) and Keck telescope. The systemic velocity of Palomar~14 is ($72.28~\pm~0.12$)~\mbox{km s$^{-1}$}. The derived velocity dispersion of $(0.38\pm0.12$)~\mbox{km s$^{-1}$} of the 16 definite member stars is in agreement with the theoretical prediction for the classical Newtonian case according to \citet{baumgardt05}. In order to exclude the possibility that a peculiar mass function might have influenced our measurements, we derived the cluster's main sequence mass function down to 0.53~$M_\odot$ using archival images obtained with the Hubble Space Telescope. We found a mass function slope of $\alpha=1.27\pm0.44$, which is, compared to the canonical mass function, a significantly shallower slope. The derived lower limit on the cluster's mass is higher than the theoretically predicted mass in case of MOND. Our data are consistent with a central density of $\rho_{0}=0.1$~\mbox{$M_\odot pc^{-3}$}. We need no dark matter in Palomar~14. If the cluster is on a circular orbit, our spectroscopic and photometric results argue against MOND, unless this cluster experienced significant mass loss. 
\end{abstract}

\keywords{globular clusters: individual (Pal~14) -- gravitation -- stellar dynamics}

\section{Introduction}
\label{intro}
Is classical Newtonian dynamics valid on all scales? On Earth classical Newtonian dynamics describes all non-relativistic phenomena very well. With the exploration and study of the Universe, we can explore low acceleration regimes that can not be studied in our backyard and we observe deviations from the expected classical Newtonian behavior. E.g. the rotation curves of spiral galaxies do not show the classically expected Keplerian fall-off, but stay flat in the outer parts of these galaxies \citep{sofue01}. These flat rotation curves are commonly explained by introducing dark matter (DM). In the outer parts of the galaxies, DM is more abundant than regular baryonic matter and the gravitational effect of the DM on the baryons results in a flat rotation curve \citep{rubin82}. A major problem DM theory has encountered recently is the discovery that young tidal-dwarf galaxies also have rotation curves that imply a significant invisible matter component although they cannot be dominated by non-baryonic DM suggesting a non-classical physical solution \citep{gentile07}.

An alternative theory to DM is modified Newtonian dynamics \citep[MOND;][]{milgrom83a, milgrom83b, bekenstein84}. In MOND, the flat rotation curves of galaxies can be fitted without any assumption of unseen matter. According to MOND, Newtonian dynamics breaks down for accelerations lower than $a_0\simeq 1\times10^{-8}$ \mbox{cm s$^{-2}$} \citep{begeman91,sanders02}. The acceleration $\textbf{\emph a}$ in \mbox{MONDian} dynamics is given by the (heuristic) equation:
\begin{equation}
\mu\left(\frac{|\textbf{\emph{a}}|}{a_0}\right)\textbf{\emph{a}}=\textbf{\emph{a}}_N,
\end{equation}
where $\mu(x)$ is an arbitrary function with the following limits:
\begin{equation}
\mu\left(x\right)=\left\{
\begin{array}{lll}
x & \mbox{if} & x\ll 1 \\
1 & \mbox{if} & x\gg 1.\\
\end{array}
\right.
\end{equation}
Here, \textbf{\emph{a}}$_N$ is the standard Newtonian acceleration and \emph{a}$_0$ is the acceleration limit below which MOND is applicable. 

It has been claimed that MOND has difficulties explaining the merging of galaxy clusters, where the baryonic matter is clearly separated from the gravitational mass, as found by gravitational lensing \citep{clowe06}. However, \citet{angus06,angus07} demonstrated that such systems are consistent with MOND, but do require the existence of some hot dark matter. 

\citet[BGK05]{baumgardt05} proposed to use distant, outer halo globular clusters (GCs) to distinguish between classical and modified Newtonian dynamics. They calculated the line-of-sight velocity dispersion for 8 Galactic GCs in classical and in modified Newtonian dynamics. For these GCs the external acceleration due to the Milky Way and the internal acceleration due to the stars themselves are significantly below the critical limit of \emph{a}$_0$. The expected velocity dispersions in case of MOND exceed those expected in the classical Newtonian case by up to a factor of three (see Table~1 in BGK05). 

Palomar~14 (Pal~14) is a diffuse GC in the outer halo of our Galaxy. Pal~14's sparseness, faintness, and large distance to the Sun makes it a difficult observational target, and therefore it did not receive much attention. The first radial velocity for a Pal~14 member star was measured by \citet{hartwick78} resulting in 81~$\pm$~3~km~s$^{-1}$. \citet{armandroff92}, based on radial velocity measurements for two stars, reported a systemic velocity of 72 $\pm$ 3 km s$^{-1}$. The  deepest ground-based color-magnitude diagram (CMD) of Pal~14 was published by \citet{sarajedini97}. He concluded that Pal~14 is 3-4~Gyr younger than halo GCs with a similar metallicity. \citet[][H06]{hilker06} published photometric data on three GCs from the BGK05 sample: AM~1, Pal~3, and Pal~14. H06 confirmed Pal~14's youth of $\sim$~10~Gyr. The data from his study are used here to obtain targets for our spectroscopic observations. \citet{dotter08} published a photometric study of Pal~14 based on archival data obtained with the Wide Field Planetary Camera 2 on board the Hubble Space Telescope. The authors confirm Pal~14's relative youth. Here the same data are used to obtain the cluster's mass function.

This paper is the second in a series that investigates theoretically and observationally the dynamics of distant star clusters. In the first paper \citep[][HBK09]{haghi09}, we derived theoretical models for pressure-supported stellar systems in general and made predictions for the outer-halo globular cluster Pal~14. In the current paper, we present a spectroscopic and photometric study of Pal~14, as a test case for the validity of MOND. I.e. we are measuring the velocity dispersion of Pal~14 in order to compare the measured value to the predicted values made for MOND and classical dynamics by HBK09. Further, we are determining the mass function of Pal~14 in order to infer the cluster's mass. The derived mass and velocity dispersion are then compared to the predictions made by HBK09 for Pal~14 on a circular orbit in MOND.

The paper is organized as follows: In Section~\ref{observations} we describe the observational material. In Sections~\ref{resultsspec} \& \ref{resultsphot} we present stellar radial velocities, the color-magnitude diagram and the mass function of Pal~14. In Section~\ref{discussion} we discuss the effects of our result for MOND and classical Newtonian gravity. The last Section concludes the paper with a summary.

\section{Observations and Data Reduction}
\label{observations}

To distinguish between MOND and classical Newtonian dynamics we used two different kinds of observations. In order to measure Pal~14's velocity dispersion we obtained high-resolution spectra of red giant candidates towards Pal~14 with the Ultraviolet-Visual Echelle Spectrograph \citep[UVES;][]{dekker00} at the Very Large Telescope (VLT) of the European Southern Observatory (ESO) in Chile and with the High Resolution Echelle Spectrograph on the Keck I telescope. To be able to measure the cluster's mass function we used imaging data from the Hubble Space Telescope archive. In the following subsections we describe the reduction process of our observational data.

\subsection{Spectroscopy with UVES}
\label{uves}

\begin{figure}
\epsscale{1.0}
\plotone{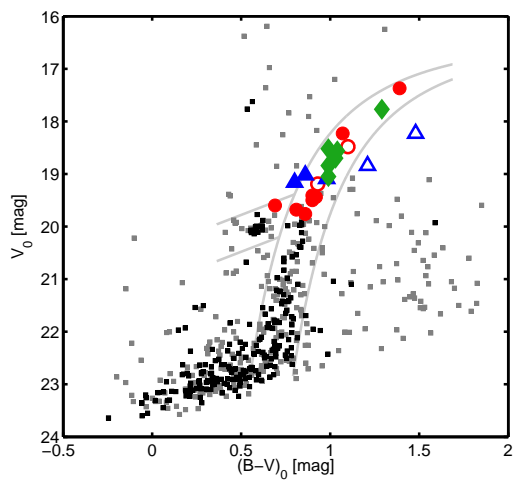}
\caption{Color-magnitude diagram of Pal~14 from \citet{hilker06}. The observed targets for the radial velocity measurements with UVES are drawn as filled and open (red) circles. The open circles denote stars that were subsequently found to be the non-members (according to their radial velocity). The (blue) filled and open triangles are the stars observed with HIRES, the open triangles are the non-members. The (green) diamonds are the stars observed with UVES and HIRES. Dark grey dots are stars within 1 half light radius of Pal~14. \label{figcmdh}}
\end{figure}

The photometry published by H06 shows the red giant branch and horizontal branch of Pal~14. Based on this photometry, we selected 16 of the 17 brightest red giants of Pal~14 for spectroscopy with UVES at the VLT. Our target stars cover the magnitude range V~=~17.3~-~19.6~mag, which includes the brightest red giant of Pal~14 and goes down to the limit of faint stars observable with UVES. Figure~\ref{figcmdh} shows Pal~14's color-magnitude diagram. 15 of our targeted stars are probable red giants and one of the targets may be an AGB or evolved horizontal branch star. The significance of this different evolutionary state will be discussed in Section~\ref{rv}. 

\begin{figure}
\epsscale{1.0}
\plotone{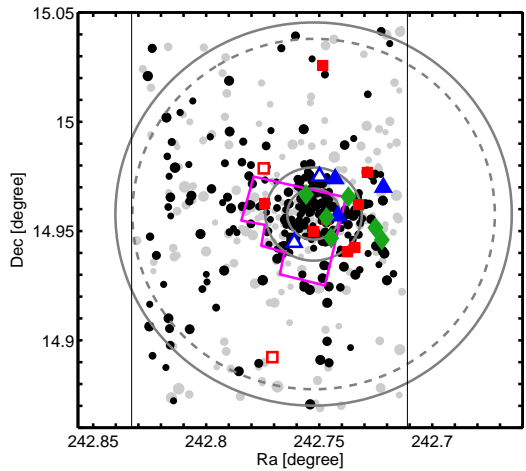}
\caption{Spatial distribution of the spectroscopically observed Pal~14 stars. The observed targets for the radial velocity measurements with UVES are drawn as (red) filled and open circles. The two open squares mark the non-members (according to their radial velocity). The (blue) open and filled  triangles are the stars observed with HIRES. The open triangles are the non-members. The (green) diamonds are the stars observed with UVES and HIRES. 
Dark grey dots are probable member stars according to their position in the CMD (see Figure~\ref{figcmdh}). The grey concentric, solid circles are from inside out the core radius, the half-light radius and the tidal radius (H06); the dashed circle is the tidal radius calculated by BGK05. The (magenta) area is the HST/WFPC2 coverage. \label{figradec}}
\end{figure}

The spatial distribution of our spectroscopic targets is shown in Figure~\ref{figradec}. The targeted stars lie mainly within two core radii with two stars in the cluster's outer region.  

We used the Besan\c{c}on Galaxy model \citep{robin03} to estimate the number of foreground stars in our sample. We extracted stars towards Pal~14 in an area on the sky covering $\sim20r_h$, where $r_h=1.28\arcmin$ is the half-light radius of Pal~14 (H06). The area covered with our sample stars is $\sim2r_h$. We selected only those stars located in the gray curves shown in Figure~\ref{figcmdh} and having apparent magnitudes V~$<20$~mag, and colors (B-V)~$>0.65$~mag. The resulting number of foreground contaminants in the actual area covered predicted by the model is $\sim1$.

UVES was used in its RED 580nm setting covering the wavelength ranges 476~-~577~nm (in the lower chip) and 584~-~648~nm (in the upper chip). We divided the 16 target stars into three setups according to their brightness: the \emph{bright} setup, containing the five brightest stars in the magnitude range m$_V$~=~17.37~-~18.52, was observed for $4\times 60$~min in total. The \emph{medium} setup with the four next fainter stars (m$_V$ = 18.56 - 19.05) was observed for $6\times 60$~min in total The \emph{faint} setup ,which included the seven least luminous stars (m$_V$ = 19.19 - 19.76), was observed for $11\times 60$~min in total. The observations were carried out in service mode within two observation periods, between May 30, 2006, and March 27, 2007. The pipeline reduced spectra (R~=~60\,000) were used for the subsequent analysis.

The zero points in the reduced spectra were not identical. The sky emission lines in the single 1h-exposures were shifted with respect to each other. To correct for this we shifted the spectra to a common position of the sky emission lines. As a sky zero point location we used the sky lines in one of our own observed sky spectra, which we defined as reference spectrum. The resulting, shifted science exposures were further corrected for the heliocentric velocity shift. Finally all the shifted single 1h-exposure spectra were co-added for each star. In this way we get for the brightest star a $S/N=16$ and for the faintest star $S/N=4$. 

\subsection{Spectroscopy with HIRES}
Within a program to study the internal kinematics of outer halo GCs \citep[for details of the program see][]{cote02} spectra for 11 candidate red giants in the direction of Pal 14 were obtained using the High Resolution Echelle Spectrograph (HIRES; Vogt et al. 1994) mounted on the Keck I telescope. The spectra, which were collected during a single night in May 1998, have a resolution of R = 45\,000 (for the 0.866\arcsec~entrance slit) and cover the wavelength range between 506~-~530~nm. The program stars were selected from CMDs published by \citet{harris84} and \citet{holland92}. The exposure times were adjusted  on a star-to-star basis, varying between 900s and 2400s with a median value of 1800s. The spectra were reduced entirely within the IRAF\footnote{IRAF is distributed by the National Optical Astronomy Observatory, which is operated by the Association of Universities for Research in Astronomy, Inc., under cooperative agreement with the National Science Foundation.}  environment, in a manner identical to that described in \citet{cote02}.

\subsection{Radial velocity}
\label{rvobs}
To measure the radial velocity of our targeted stars we cross-correlated our final UVES and HIRES spectra with two high-resolution spectra of the UVES Paranal Observatory Project \citep[UVES POP;][]{bagnulo03}: HD37811 (a G7 red giant) and HD45415 (a G9 red giant). The cross-correlation was done with the IRAF task \emph{fxcor}. The heliocentric radial velocities of our two standard stars are: $v_{HD37811}$ = (-4.68$\pm$0.11)~\mbox{km s$^{-1}$}, $v_{HD45415}$ = (52.70$\pm$0.04)~\mbox{km s$^{-1}$} (Melo 2007, private communication).

We determined the velocity shift of our sample stars relative to each of the two UVES POP stars. The UVES camera consists of two CCDs. For each pair of a UVES science target star and of a UVES POP star, we determined two radial velocities, one for the upper UVES chip and one for the lower UVES chip. These two velocities are averaged to a final velocity relative to the UVES POP star. Comparing the relative velocities measured for the two UVES POP stars, we find a mean difference of 0.1~\mbox{km s$^{-1}$}. Within the errors the two velocities are equal. The UVES science stars' radial velocity is the mean of the two velocities weighted by the Tonry-Davis R value \citep{tonry79} determined by \emph{fxcor}. 

For the HIRES sample, we determined the velocity of each science star relative to both UVES POP stars. Comparing the two relative velocities we find a mean difference of 0.07~\mbox{km s$^{-1}$}. The HIRES stars' radial velocity is the mean of these two measured velocities weighted by the Tonry-Davis R value. 

For 6 stars we have both UVES and HIRES spectra. To determine a common zero point of the two different samples we compared the measured velocities for these 6 stars. A mean velocity shift of $\Delta v$ = 0.64~\mbox{km s$^{-1}$} was found. The shift is probably due to a different instrumental zero point. The final HIRES velocities are corrected for this shift. The shift can also be due to binarity or stellar variability. For two of the six stars we also have UVES measurements at two epochs, within the errors the velocities agree very well. Short-period binarity and variability can be excluded for these two stars. The error of the HIRES measurements for the five fainter stars is comparable to the mean shift. Five stars have a positive velocity shift and only one a negative. If all stars were binaries we would not expect a clear spread around a positive shift.

The final radial velocity for the 6 stars, with UVES and HIRES spectra, is the weighted mean of the measured velocities. For the remaining 15 stars we only have measurements of one instrument, therefore this velocity is taken as the final radial velocity of the star.

\subsection{Photometry}
\label{phot}
We used imaging data obtained with the Hubble Space Telescope/Wide Field Planetary Camera 2 (HST/WFPC2) from the HST archive to obtain a deep CMD of Pal~14. The data were obtained as part of the proposal GO-6512 (PI: Hesser). The same data were used by \citet{dotter08}. The WFPC2 images cover the entire area within the cluster's core radius (H06), about $67\%$ of the area within the nominal half-light radius (H06), and only $7\%$ of the area within the tidal radius (H06) (see Figure~\ref{figradec}). The pipeline-reduced FITS files were run through multidrizzle/tweakshifts \citep{koekemoer02} to refine the image registration. All further processing was done on the original files together with the refined shifts, using the WFPC2 photometry package HSTphot \citep{dolphin00} and following the strategy outlined in the HSTphot User's Guide for preprocessing, photometry and artificial star tests. As in each subset of well aligned images in the same filter, the exposure times differed significantly, no co-adding was done. In Figure~\ref{figcmdh} we show the CMD of all stars brighter than 28$^{th}$~mag detected by HSTphot with the following selections: the HSTphot sharpness parameter ($|sharpness|<0.2$), HSTphot type parameter ($type<3$, i.e. the star is either a single star or a possible unresolved binary), and magnitude errors $\sigma_{mag}<0.2$~mag. 

\begin{figure}
\epsscale{1.0}
\plotone{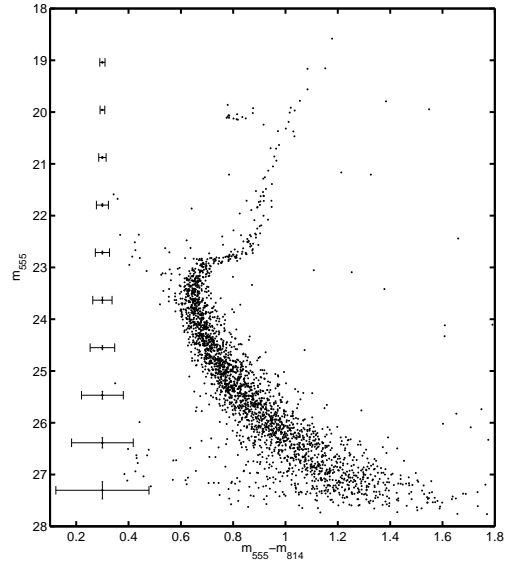}
\caption{Color-magnitude diagram of Pal~14. We show the remaining stars after applying selections in the HSTphot parameters: sharpness, magnitude errors, and type. The CMD contains 2752 stars.\label{figcmds}}
\end{figure}

To determine the photometric errors we inserted artificial stars with known magnitudes. The deviations of the subsequently measured magnitudes to the inserted values let us determine the photometric errors shown in Figure~\ref{figcmds}.

\subsection{Completeness}
\label{comp}
\begin{figure}
\epsscale{1.0}
\plotone{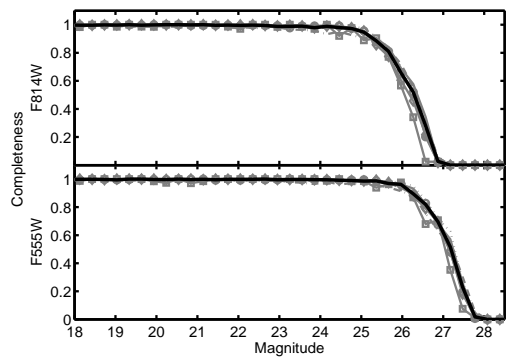}
\caption{Completeness for the two filters F555W (lower plot) and F814W (upper plot). For each filter the completeness profile for seven annuli of width 0.3~arcmin are plotted as gray lines with different symbols, the black line is the overall completeness used in the data analysis.\label{figcomp}}
\end{figure}

For a detailed analysis of the stars in Pal~14, we performed radius-dependent artificial star tests within HSTphot to determine the completeness of the observations. For the artificial star experiment we added $\sim$160\,000 stars onto the image. For 7 annuli of a width of \mbox{0.3~arcmin} we counted the number of artificial stars retrieved from the image with a magnitude not more than 0.2~mag different from the input value. In Figure~\ref{figcomp} we show the seven completeness profiles (gray curves), which essentially fall on top of each other. Therefore, no radial dependence is observed, which is mainly due to the low density of Pal~14. The profile of the outermost annulus (solid line with squares) shows a decline at slightly brighter magnitudes. This is an artificial effect. The number of stars in this annulus is only 10\% of the average number of stars in the other annulii. We used an averaged completeness profile in our analysis, shown as the black line. The $50\%$ completeness limit is reached at $m_{555}$ = $27.21$~mag.

\section{Spectroscopic Results}
\label{resultsspec}

\subsection{Individual stellar radial velocities}
\label{rv}

\begin{deluxetable}{rrccccccccccc}
\tablewidth{0pt}
\tablecolumns{13}
\tablecaption{Heliocentric radial velocities of our sample stars\label{tblvel}}
\tabletypesize{\scriptsize}
\tablehead{
\colhead{Star\tablenotemark{a}} &\colhead{Star\tablenotemark{b}} & 
\colhead{$\alpha$(2000)}        &\colhead{$\delta$(2000)}        &
\colhead{$m_V$}                 &\colhead{$B-V$}                 &
\colhead{\emph{v(UVES)}}        &\colhead{$\sigma_{v(UVES)}$}    &
\colhead{\emph{v(HIRES)}}       &\colhead{$\sigma_{v(HIRES)}$}   & 
\colhead{$v_{rad}$}             &\colhead{$\sigma_{v_{rad}}$}    & 
\colhead{m?}\\
\colhead{}                      &\colhead{}                      &
\colhead{hh:mm:ss.ss}           &\colhead{$^\circ$:':''}         &
\colhead{mag}                   &\colhead{mag}                   & 
\colhead{$km~s^{-1}$}           &\colhead{$km~s^{-1}$}           &
\colhead{$km~s^{-1}$}           &\colhead{$km~s^{-1}$}           &
\colhead{$km~s^{-1}$}           &\colhead{$km~s^{-1}$}           &
\colhead{}}
\startdata
1 &      & 16:11:05.81 &14:57:45.1 & 17.37&1.39 &  72.53& 0.07  &\nodata&\nodata&72.53  &0.07   & Y\\
2 & HV025& 16:10:58.73 &14:56:48.7 & 17.77&1.29 &  72.76& 0.09  &  71.49& 0.30  &72.47  &0.14   & Y\\
3 &      & 16:10:54.90 &14:58:36.7 & 18.23&1.07 &  71.75& 0.14  &\nodata&\nodata&71.75  &0.14   & Y\\
  & HV051& 16:10:59.98 &14:58:30.1 & 18.23&1.48 &\nodata&\nodata&-73.77 & 1.17  &\nodata&\nodata& N\\
4 &      & 16:11:04.98 &14:53:32.3 & 18.48&1.10 & -32.14& 0.16  &\nodata&\nodata&\nodata&\nodata& N\\
5 & HV007& 16:10:59.24 &14:57:22.5 & 18.52&0.99 &  71.68& 0.18  &  73.23& 0.53  &72.21  &0.30   & Y\\
6 & HH244& 16:10:53.36 &14:56:45.4 & 18.56&1.04 &  72.58& 0.18  &  72.79& 0.46  &72.65  &0.27   & Y\\
7 & HH201& 16:10:54.04 &14:57:05.6 & 18.70&1.03 &  72.62& 0.18  &  72.68& 0.49  &72.64  &0.27   & Y\\
8 & HV043& 16:10:56.90 &14:57:56.5 & 18.84&0.99 &  71.56& 0.21  &  70.97& 0.47  &71.38  &0.31   & Y\\
  & HV086& 16:11:02.66 &14:56:41.1 & 18.84&1.21 &\nodata&\nodata&-155.31& 0.85  &\nodata&\nodata& N\\
  & HV055& 16:10:58.31 &14:58:26.2 & 19.02&0.86 &\nodata&\nodata&  73.62& 0.89  &73.62  &0.89   & Y\\
9 & HV104& 16:11:01.40 &14:57:60.0 & 19.05&0.99 &  73.49& 0.21  &  73.53& 0.91  &73.50  &0.43   & Y\\
  & HH042& 16:10:53.20 &14:58:12.0 & 19.09&0.98 &\nodata&\nodata&  71.94& 0.35  &71.94  &0.35   & Y\\
  & HV004& 16:10:58.03 &14:57:25.1 & 19.16&0.80 &\nodata&\nodata&  73.23& 0.56  &73.23  &0.56   & Y\\
10&      & 16:11:05.89 &14:58:43.2 & 19.19&0.93 &  50.44& 0.19  &\nodata&\nodata&\nodata&\nodata& N\\
12& HV074& 16:10:56.21 &14:56:32.7 & 19.41&0.90 &  71.83& 0.23  &\nodata&\nodata&71.83  &0.23   & Y\\
13& HV075& 16:10:56.98 &14:56:25.8 & 19.44&0.92 &  72.33& 0.41  &\nodata&\nodata&72.33  &0.41   & Y\\
14& HV006& 16:10:59.24 &14:57:19.7 & 19.50&0.90 &  71.80& 0.27  &\nodata&\nodata&71.80  &0.27   & Y\\
15& HV042& 16:10:55.84 &14:57:43.4 & 19.60&0.69 &  69.99& 0.38  &\nodata&\nodata&69.99  &0.38   & Y?\\
16&      & 16:10:59.62 &15:01:32.9 & 19.68&0.81 &  72.14& 0.43  &\nodata&\nodata&72.14  &0.43   & Y\\
17& HV021& 16:11:00.58 &14:56:59.1 & 19.76&0.86 &  72.39& 0.32  &\nodata&\nodata&72.39  &0.32   & Y\\
\enddata
\tablenotetext{a}{Identification from \citet{hilker06}}
\tablenotetext{b}{Identification from \citet{harris84} and \citet{holland92}}
\end{deluxetable}

In Table~\ref{tblvel}, we list the measured heliocentric radial velocities (\emph{v(UVES)} and \emph{v(HIRES)}) and their errors ($\sigma_{v(UVES)}$ and $\sigma_{v(HIRES)}$) for the 21 stars in our sample. The listed velocities $v_{rad}$ are the weighted mean of the UVES and the HIRES observations. The listed errors are the propagated errors, weighted by the Tonry-Davis R value from the cross-correlations. Star~4, Star~10, HV051 and HV086 all have significantly different velocities than the majority of the measured stars which are centered around $\sim72.2$~\mbox{km s$^{-1}$}: $v_{Star 4}=(-32.14\pm0.16)$~\mbox{km s$^{-1}$}, $v_{Star 10}=(50.44\pm0.18)$~\mbox{km s$^{-1}$}, $v_{HV051}=(-74.41\pm1.17)$~\mbox{km s$^{-1}$} and $v_{HV086}=(-155.95\pm0.85)$~\mbox{km s$^{-1}$}.  These four stars are categorized as non-members (open circles and triangles in Figures~\ref{figcmdh} \&~\ref{figradec}). The remaining 17 stars are considered to be members of Pal~14 (see last column in Table~\ref{tblvel}). The measured velocity of Star~15 is more than $3\sigma$ away from the mean of the other member stars. Therefore, we present all our results including and excluding Star~15.

\subsection{The systemic velocity and the velocity dispersion}

\begin{figure}
\epsscale{1.0}
\plotone{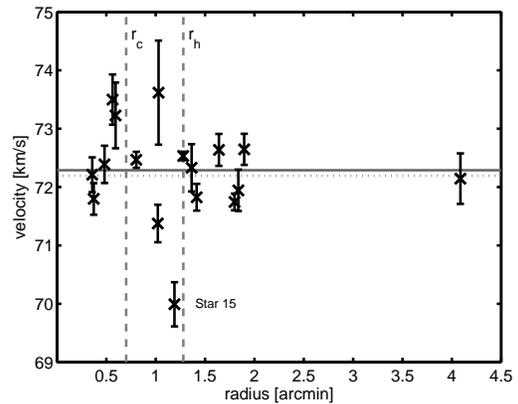}
\caption{Radial distribution of stars with velocity measurements in Table~\ref{tblvel}. Black crosses indicate the 17 member stars of our sample of Pal~14. Star~15, for which the measured velocity is suspicious, is labeled. The horizontal solid line marks Pal~14's global radial velocity without Star~15, and the dotted line the radial velocity including Star~15. Further the core and half-light radii are indicated by dashed grey lines for an easier comparison with Figure~\ref{figdisp}. \label{figvel}}
\end{figure}

\begin{deluxetable}{ccccc}
\tablewidth{0pt}
\tabletypesize{\scriptsize}
\tablecaption{Radial velocity and velocity dispersion of Pal~14\label{tblvelcomb}}
\tablehead{
\colhead{Instrument} & \colhead{$velocity$} & \colhead{$error_{v}$}& \colhead{$dispersion$} & \colhead{$error_{d}$}\\
\colhead{} &  \colhead{$km~s^{-1}$} & \colhead{$km~s^{-1}$}& \colhead{$km~s^{-1}$} & \colhead{$km~s^{-1}$}
}
\startdata
UVES\tablenotemark{a} & 72.28 (72.12) & 0.15 (0.20) & 0.50 (0.70) & 0.11 (0.15)\\
HIRES\tablenotemark{a} & 72.46& 0.29&0.66& 0.26\\
combined & 72.28 (72.19)& 0.12 (0.18)& 0.38 (0.64)& 0.12 (0.15)
\enddata			     

\tablenotetext{a}{The first value is without the measurement of Star~15. The value in parentheses includes the measurement of Star~15.}
\end{deluxetable}
First, we determined the mean velocity and the global velocity dispersion for the two different measurement sets, respectively. We used the maximization method described in \citet{pryor93}. The mean velocity for the HIRES measurements is ($72.46~\pm~0.29$)~\mbox{km s$^{-1}$} and the velocity dispersion ($0.66~\pm~0.26$)~\mbox{km s$^{-1}$}. For the UVES measurements we find a mean velocity of ($72.28~\pm~0.15$)~\mbox{km s$^{-1}$} and a velocity dispersion of ($0.50~\pm~0.11$)~\mbox{km s$^{-1}$} if we exclude Star~15. Including Star~15, we find ($72.12~\pm~0.20$)~\mbox{km s$^{-1}$} and ($0.70~\pm~0.15$)~\mbox{km s$^{-1}$}. The measurements of the two samples agree very well.

Second, to determine the overall mean velocity and the global dispersion for all stars we also used the maximization method of \citet{pryor93}. Including Star~15, we measured a mean heliocentric radial velocity for Pal~14 of ($72.19~\pm~0.18$)~\mbox{km s$^{-1}$}, excluding Star~15, ($72.28~\pm~0.12$)~\mbox{km s$^{-1}$}. Within the error bars the two values agree. Our results confirm the earlier measurements by \citet{armandroff92}. 

Figure~\ref{figvel} shows the radial profile of our measured velocities (Star~15 is labeled). The cluster's mean velocity (for both cases) is marked by the solid (without Star~15) and dotted (with Star~15) horizontal line. In Table~\ref{tblvelcomb} we summarize the radial velocity and velocity dispersion measurements for the two instruments and for the combined stellar sample.

The global line-of-sight velocity dispersion for Pal~14 with Star~15 included is ($0.64~\pm~0.15$)~\mbox{km s$^{-1}$} with $99\%$ confidence limits of $0.41$~\mbox{km s$^{-1}$} and $1.10$~\mbox{km s$^{-1}$}. Without Star~15, the line-of-sight velocity dispersion is ($0.38~\pm~0.12$)~\mbox{km s$^{-1}$} with $99\%$ confidence limits of $0.26$~\mbox{km s$^{-1}$}  and $0.67$~\mbox{km s$^{-1}$}. Within the errors the two values would agree. The theoretical prediction for the velocity dispersion of BGK05, for which a $M/L=2$ was assumed, is $\sigma_{MOND}=1.27$~\mbox{km s$^{-1}$} and $\sigma_{Newton}=0.52$~\mbox{km s$^{-1}$}. For both cases, when Star~15 is included or excluded, our results are more consistent with the classical Newtonian prediction, while the \mbox{MONDian} prediction is outside the $99\%$ confidence limits. 

\begin{figure}
\epsscale{1.0}
\plotone{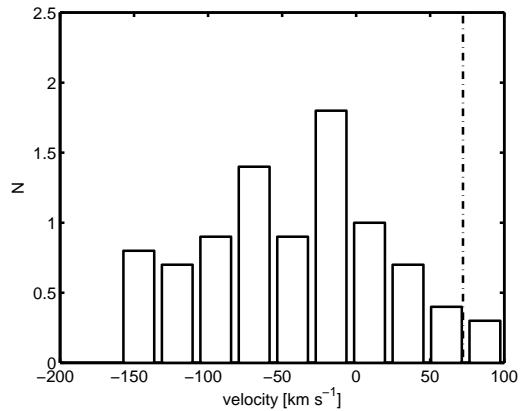}
\caption{Expected velocity distribution based on the Besan\c{c}on Galaxy model. Only stars that lie within the grey area in Figure~\ref{figcmdh} and that have a radial velocity larger than $-160$~\mbox{km s$^{-1}$} are counted (see text for more details). The dash-dotted, vertical line marks the systemic velocity of Pal~14, $\sim72.2$~\mbox{km s$^{-1}$}.\label{figveldispexp}}
\end{figure}

As described above the measured velocity of Star~15 seems to be deviant. There are several possible explanations for this discrepant velocity of Star~15: 
\emph{i)}~Star~15 is a normal member of Pal~14. We performed a Monte Carlo simulation in order to evaluate how likely the measured radial velocity profile is. In the Monte Carlo simulation we randomly drew velocities from a Gaussian distribution, which was newly initialized for each draw by calculating the mean velocity and standard deviation of our measured radial velocities randomly convolved with their errors. The radial distributions of all draws were added. We performed a KS-test of the simulated velocity distribution with the distribution of the actually measured velocity. The KS-test revealed a $<1\%$ probability that the distribution that includes Star~15 comes from a Gaussian distribution, whereas the probability was $\sim50\%$ that the distribution without Star~15 is Gaussian. This argues against Star~15's membership in Pal~14.
\emph{ii)}~Star~15 is not a red giant, but more likely an evolved horizontal branch (HB) star or an AGB star judging from its position in the CMD (see Figure~\ref{figcmdh}). The used templates of a G7 (HD37811) and a G9 (HD45415) red giant may not be appropriate for Star~15.
\emph{iii)}~Star~15 could be a binary. For our faint UVES sample (Stars~10-17) we have observations at two epochs: June, 2006 and March, 2007. Theoretically this allows us to measure a possible change in velocity due to binarity. The faintness and the therefore low $S/N$ of Star~15's spectra does not allow us to accurately measure the individual radial velocity for both epochs. The two measured velocities are $v_{2006}=(70.64~\pm~0.63)$~\mbox{km s$^{-1}$} and $v_{2007}=(69.13~\pm~0.75)$~\mbox{km s$^{-1}$}. Within the errors the two velocities are the same. Nonetheless, this does not allow us to exclude long-period binarity.
\emph{iv)}~A further cause for the large offset of Star~15's velocity could be strong atmospheric variability, which can occur among AGB stars. However, from its position in the CMD, Star~15 would be an early-AGB star. In this early phase, AGB stars are not yet pulsating very strongly \citep{habing03}. With essentially only one observing epoch it is impossible to know about the star's variability.
\emph{v)}~Another option might be that Star~15 is not a member of Pal~14. We computed a model velocity distribution of stars which are located within the light gray curves shown in Figure~\ref{figcmdh} using the Besan\c{c}on Galaxy model \citep{robin03} as described in Section~\ref{uves}. The expected velocity distribution, for stars with radial velocities $>-160$~~\mbox{km s$^{-1}$}, is shown in Figure~\ref{figveldispexp}. The number of stars in each bin is scaled to an area of $\sim2r_h$, in order to reproduce the actually observed area. We expect about 9 stars to fulfill the photometric constraints. 0.5 stars have a radial velocity between 50~\mbox{km s$^{-1}$} and 75~\mbox{km s$^{-1}$}. Therefore, Star~15 could be a foreground contaminant.

\section{Photometric results}
\label{resultsphot}
In order to make predictions for the velocity dispersion in Newtonian and \mbox{MONDian} dynamics, we first have to determine the mass of Pal~14. The measured low velocity dispersion is in excellent agreement with the theoretical prediction of classical Newtonian dynamics, and a very strong indicator against MOND. The theoretical calculations by HBK09 show the dependence of velocity dispersion and mass (see Figure~8 in HBK09) for classical dynamics and MOND. For a given velocity dispersion the necessary mass is always smaller in MOND than in classical dynamics. Our derived low velocity dispersion is explainable in MOND if we find a low total mass for Pal~14. To constrain the mass in Pal~14, we analyzed Pal~14's CMD and main-sequence mass function.

\subsection{Color-magnitude diagram}
\begin{figure}
\epsscale{1.0}
\plotone{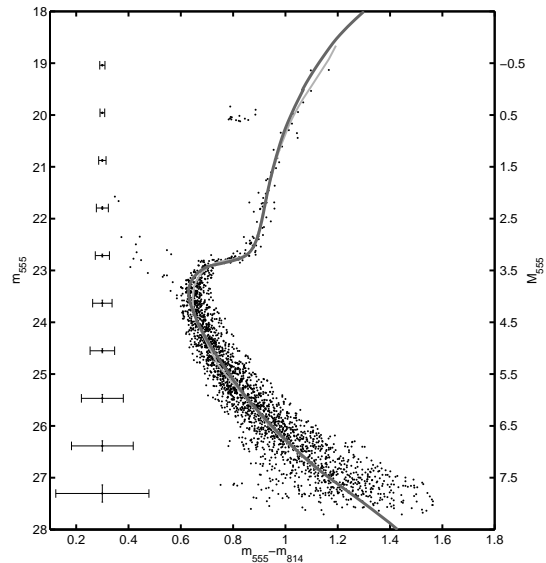}
\caption{Color-magnitude diagram of our final sample of Pal~14 stars from WFPC2. The light gray line shows the derived cluster ridgeline, the dark gray line is the best fitted $\alpha$-enhanced, [$\alpha$/Fe]=+0.2, Dartmouth isochrone with an age of 11.5~Gyr, E(m$_{555}$-m$_{814}$) = 0.063, and $(m-M)_{555}$~=~19.45~mag. \label{figcmdage}}
\end{figure}

Figure~\ref{figcmds} shows the HST CMD of Pal~14 with the remaining stars after the HSTphot parameter cuts (see Section~\ref{phot} for details). The CMD reaches $\sim$4 mag below the main-sequence turnoff, m$_{MSTO}$~=~23.63~$\pm$~0.01~mag, which allows us to theoretically determine the cluster's mass function down to $\sim$0.49~$M_{\odot}$ (see Section~\ref{mass}). The CMD shows a well-populated main-sequence (MS), subgiant branch, red giant branch, red horizontal branch (HB), and some probable blue straggler stars. The presence of a red HB and its implications were discussed in \citet{sarajedini97} and \citet{dotter08}. As expected there is only little field star contamination of Milky Way stars due to the moderately high Galactic latitude of Pal~14 and due to the small field of view of WFPC2. Judging from the TRILEGAL Galaxy Model \citep{girardi05}, the number of contaminating foreground stars on our WFPC2 image in the CMD-area covered by Pal~14 is $\sim 2$. The width of the main-sequence which we observe is due to a combination of the photometric errors and binary stars.

For our further analysis of Pal~14, we applied a stricter selection of our stellar sample. We determined the cluster's fiducial ridgeline (see Figure~\ref{figcmdage}, the light gray line). The ridgeline reproduces the mean location of the stellar distribution in the CMD. To derive the cluster's ridgeline we adopted the method described in \citet{glatt08a}. We selected all stars within 2$\sigma$ of the ridgeline and added the blue stragglers and the HB stars for our final sample. The 2\,500 stars in our final sample are plotted in Figure~\ref{figcmdage}.

\subsection{Age \& Distance}
Pal~14 is known to be younger than typical halo GCs \citep{sarajedini97,hilker06,dotter08} at its metallicity. We derived Pal~14's age via isochrone fitting. We used the Dartmouth isochrones \citep{dotter07}, which have been shown to reproduce the location of the MS, subgiant branch, and red giant branch very well \citep{glatt08b}. We adopted the published spectroscopically determined metallicity of \mbox{[Fe/H]~=~-1.50} \citep{harris96}. Distance and reddening were treated as free parameters. A large number of isochrones was fitted using different combinations of age, distance, and reddening. We selected by trial-and-error the isochrone that best matched the above derived ridgeline. 

With an $\alpha$-enhanced isochrone, [$\alpha$/Fe]~=~+0.2, our best fit yields an age of (11.5$\pm$0.5)~Gyr, a reddening of E${(m_{555}-m_{814})}$ = 0.063 (corresponding to E(B-V)~=~E${(m_{555}-m_{814})}$/1.2~=~0.05 \citep{holtzman95}), and an extinction corrected distance modulus of \mbox{$(m-M)_{555,0}$~=~19.25~mag}.  
\citet{sarajedini97} stated the age of Pal~14 is 3-4~Gyr younger than the age of similar halo GCs, H06 derived an age of 10~Gyr and \citet{dotter08} determined an age of 10.5 Gyr via $\alpha$-enhanced isochrone fitting. An $\alpha$-enhancement is found for many of the Milky Way GCs \citep[see, e.g.][]{carney96}. Our new age determination reduces the offset to other halo GCs slightly.

From our CMD and the isochrone fit, we find a dereddened distance to Pal~14 of ($71\pm1.3$)~kpc, which places Pal~14 a bit closer to the Sun than previously thought. In comparison, H06 derived a distance to Pal~14 of 74.7~kpc. \citet{dotter08} derived an even larger distance of 79~kpc. 

\subsection{Luminosity function}
\label{lum}

The cluster's MS luminosity function was derived by counting the number of stars, fainter than the MS turnoff at m$_{MSTO,0}$ = 23.44$\pm$0.01~mag, in 0.5~mag wide bins separated by 0.1~mag along the dereddened m$_{555}$ axis.

\begin{figure}
\epsscale{1.0}
\plotone{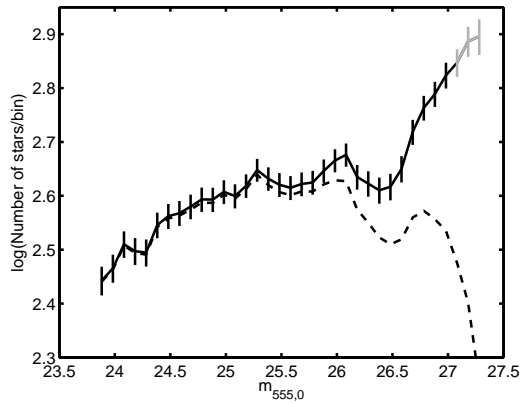}
\caption{Luminosity function of Pal~14's MS. The dashed line is the number of observed stars corrected for the missing area coverage The solid line is the number of stars after the correction for photometric incompleteness (the grey dots mark points with a completeness $<$50\%). The horizontal bars are the $N^{1/2}$ errors.\label{figlf}}
\end{figure}

Furthermore, the WFPC2 images do not cover the entire projected spatial extension of the cluster on the sky. Our data cover the entire area within the cluster's core radius (H06), about $67\%$ of the area within the nominal half-light radius (H06), and only $7\%$ of the area within the tidal radius (H06). The correction for the missing coverage within the half-light radius was done as follows. We derived the luminosity function for the stars within the annulus between the half-light and the core radius ($n_{annulus}$). We then corrected each magnitude bin of the entire distribution proportionally to the distribution of stars within the covered annulus: 
\begin{eqnarray}
N_{area}&=&N_{obs}+n_{annulus}(\frac{A_{annulus}}{A_{covered}}-1),
\end{eqnarray}
where $A_{covered}$ is the area of the annulus covered by the WFPC2 image, and $A_{annulus}$ the area of the annulus itself. The final correction was done for the photometric incompleteness (see Figure~\ref{figcomp} and Section~\ref{comp}). We did not correct for potential foreground contaminants. The TRILEGAL Galaxy model only predicts a very small number of stars on our main-sequence. In Figure~\ref{figlf} we show the resulting luminosity function of Pal~14. The solid line shows the final number of stars per 0.5~mag bin. The errors are given as $N^{1/2}$.

\citet{dotter08} report an unusual flat luminosity function for Pal~14 between V~=23~mag and 28~mag. Their data was not corrected for incompleteness. Our MS luminosity function shows the same flat behavior, correcting for incompleteness does not change the slope dramatically.

\subsection{Mass function}
\label{mass}

The function $dN/dm \propto m^{-\alpha}$ describes the number of stars in the mass interval [$m$, $m+dm$]. We obtained such a mass function for Pal~14's MS. The upper boundary of the MS is at the turnoff, m$_{MSTO,0}$ = 23.44$\pm$0.01~mag. Using the masses given by the 11.5 Gyr isochrone by \citet{dotter07}, we have stellar masses on the MS covered by our photometry between 0.49 $M_\odot$ and 0.79~$M_\odot$. We binned the masses linearly into 10 bins of equal width of 0.03 $M_\odot$. In Table~\ref{tblmf}, we list the center of the mass bins in the first column, and the number of observed stars (N$_{obs}$) for each bin in the second column.

\begin{deluxetable}{cccccc}
\tablewidth{0pt}
\tabletypesize{\scriptsize}
\tablecaption{Mass function of Pal~14\label{tblmf}}
\tablehead{
\colhead{Bin center} & \colhead{N$_{obs}$} & \colhead{N$_{corrected}$} & \colhead{N$_{f}$} & \colhead{$\sigma_{N_f}$} &\colhead{Completeness} \\  
\colhead{$\mathcal{M}_{\odot}$}   
}
\startdata
0.51 & 114 & 152 & 706 & 59 & 0.21\\
0.54 & 249 & 330 & 542 & 30 & 0.61\\
0.57 & 196 & 255 & 316 & 20 & 0.81\\
0.60 & 258 & 328 & 372 & 21 & 0.91\\
0.63 & 201 & 274 & 282 & 17 & 0.97\\
0.66 & 226 & 306 & 311 & 18 & 0.98\\
0.69 & 219 & 282 & 286 & 17 & 0.99\\
0.72 & 212 & 271 & 273 & 17 & 0.99\\
0.75 & 213 & 268 & 270 & 17 & 0.99\\
0.78 & 225 & 301 & 302 & 18 & 1.00\\
0.81 & 289 & 366 & 367 & 20 & 1.00
\enddata       
\tablecomments{Column~1 lists the center of our mass bins, column~2 the number of observed stars per bin, column~3 the number of stars per bin after correcting for the missing area coverage, column~4 contains the final number of stars per bin after correcting for completeness, column 6 lists the propagated error on the final number of stars per bin, and column~7 lists the average completeness value for the mass bin. (The numbers in columns~3,4, and 5 are rounded to the nearest integer.)}
\end{deluxetable}

\begin{figure}
\epsscale{1.0}
\plotone{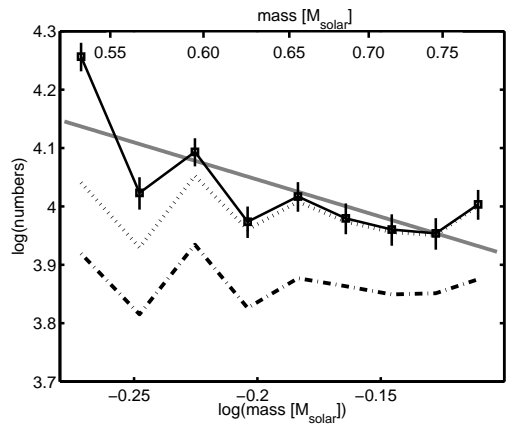}
\caption{Mass function of Pal~14. The lowest (dash-dotted) line is the observed mass function for stars with masses between 0.49 $M_\odot$ and 0.80 $M_\odot$. The dotted line shows the mass function after correcting the number of stars per bin for the missing area coverage of the WFPC2 data. The top line also includes the correction for the photometric completeness. In gray the fitted slope $\alpha = 1.27\pm0.44$ is shown.\label{figmf}}
\end{figure}

We corrected the number of stars per mass bin for the same effects as in the case for the luminosity function. First, the observed number of stars per mass bin was corrected for the missing area coverage in the same way as described above (Table~\ref{tblmf}, column~3). Second, the mean of the stars' incompleteness was used as a correction factor. The corrected number of stars per mass bin is listed in Table~\ref{tblmf}, column 4. To fit a slope to our data we only considered data with a completeness factor $> 0.50$ (see last column in Table~\ref{tblmf}). This restriction leads to a MS mass function covering the range from 0.525~$M_\odot$ to 0.79~$M_\odot$. We fitted a slope to our data points in $log(number)$ vs $log(mass)$ space. In Figure~\ref{figmf} we plot the resulting mass function and the fitted slope of $\alpha = 1.27\pm 0.44$ as the gray line. \citet{dotter08} find a similar mass function slope of $\alpha\approx1.2$. The canonical Kroupa IMF \citep{kroupa01} in this notation is $2.35$ for the given mass range. In Figure~\ref{figmf}, the observed mass function is shown as the dash-dotted line. The  dotted line is the mass function after the area corrections. The solid line denotes the completeness corrected number of stars. 

Compared to the canonical slope of $2.35$ Pal~14's mass function is flatter in the given mass range. \citet{demarchi07} compiled the mass function slope in the stellar mass range 0.3 to 0.8~$M_\odot$ for 20 Galactic GCs of different sizes, concentrations, positions in the Galaxy, etc. Pal~14 has a (measured) concentration of c~=~0.85 (H06). Clusters with a similar concentration span a mass function slope range of $\alpha = -0.9 \dots 1.3$ (see Figure~1 in \citet{demarchi07}). The derived slope is comparable with the slope of similar clusters. E.g., NGC6809 has a concentration of 0.76 and a mass function slope of 1.3. This slope was derived around the cluster's half-light radius, where the impact of mass segregation is negligible \citep{paresce00}. For Pal~14 we see an increasing number of stars per unit mass down to 0.525~$M_\odot$. A sudden decrease below this low-mass limit would be a unique case as no Galactic GC is known to show an initial rise followed by a decrease. 

In principle there are two reasons for such a depleted mass function: Either Pal~14 did form with only few low-mass stars, or the cluster is mass segregated and lost most of its low-mass stars through interaction with the Galactic tidal field. The small area covered by the WFPC2 image does not allow us to estimate the amount of mass segregation. In an upcoming paper we will discuss the issue of mass segregation in Pal~14 based on imaging data we obtained at the VLT.

\subsection{Total mass \& mass-to-light ratio}
To estimate the mass of Pal~14 we corrected for the missing area within the half-light radius. We measured an observed mass for Pal~14's main-sequence \mbox{$\mathcal{M}_{ms, obs}$ = (1\,340 $\pm$ 50)~$M_\odot$} (above the 50\% completeness limit). The errors are propagated from the measured photometry. Taking into account the stars brighter than the MS turn off, correcting for the missing area within the half-light radius and the completeness we get \mbox{$\mathcal{M}_{cor}$ = (2\,200 $\pm$ 90)~$M_\odot$} within the mass range 0.525~$M_\odot$ to 0.83~$M_\odot$. If we extrapolate by assuming that the measured slope of $\alpha =1.27$ holds down to 0.5~$M_\odot$ and assume a Kroupa-like mass function, $\alpha =1.3$  for masses between 0.1~$M_\odot$ and 0.5~$M_\odot$, we have a total mass within the half-light radius for Pal~14 of \mbox{$\mathcal{M}_{tot,hl}$ = (6\,020 $\pm$ 500)~$M_\odot$}.

The slope of the mass function for stars with masses $<0.5$~$M_\odot$ is still under debate \citep{kroupa02,elmegreen08}. Pal~14 is very far from the Milky Way. It may have an eccentric orbit that would bring it much closer to the Milky Way at perigalacticon possibly leading to strong tidal interaction and to an enhanced loss of very low-mass stars. \citet{richer04,richer08} studied the main-sequence mass function of the GCs NGC~6397 and M~4 down to the hydrogen-burning limit. In the cluster cores they found mass function slopes of \mbox{$\alpha = -0.7$}, these cluster centers lack low-mass stars. Therefore, we also calculated the mass in Pal~14 for a mass function with a linearly declining slope for masses $<0.5$~$M_\odot$ towards less massive stars (\mbox{$\alpha = -1.0$}). In that case the lower limit for the total mass within the half-light radius of Pal~14 is \mbox{$\mathcal{M}_{tot,hl}$ = (2\,930 $\pm$ 130)~$M_\odot$}. 

If we assume that light traces mass, then the half-light radius will also be the half-mass radius. Therefore we double the above numbers to estimate the total mass of Pal~14. The extrapolation with a Kroupa-like IMF for stellar masses between 0.1~$M_\odot$ and 0.5~$M_\odot$ yields a total mass of Pal~14 of $\mathcal{M}_{tot}$~$\approx$~12\,040~$M_\odot$. With the declining mass function for masses $<0.5$~$M_\odot$, we get a total mass of $\mathcal{M}_{tot}$~$\approx$~5\,860~$M_\odot$. Considering stellar remnants will increase the mass further.

Using the total mass of Pal~14, we derive the mass-to-light ratio. The extrapolation with the Kroupa-like mass function yields \mbox{$M/L$~=~($2.2\pm0.4$)~$M_\odot/L_\odot$}. The extrapolation with the declining mass function gives \mbox{$M/L$~=~($1.1\pm0.1$)~$M_\odot/L_\odot$}.

\section{Discussion}
\label{discussion}

\subsection{MOND?}
\begin{figure}
\epsscale{1.0}
\plotone{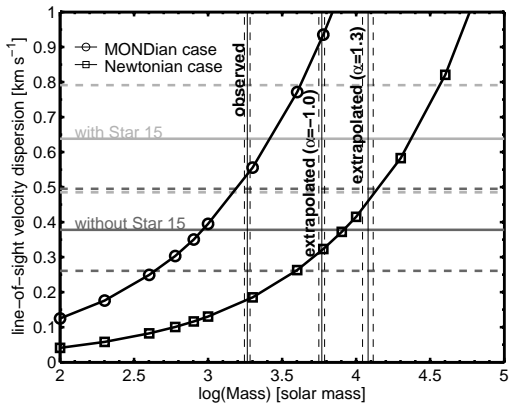}
\caption{Theoretically predicted velocity dispersion as a function of mass. The two black curves are the predictions in \mbox{MONDian} dynamics (open circles) and in classical Newtonian dynamics (open squares). The observed velocity dispersions (and the errors) are drawn as the two horizontal lines, the light gray without Star~15, dark gray with Star~15. The vertical lines mark the observed lower mass limit and the two extrapolated lower mass limits.\label{figtheory}}
\end{figure}

In HBK09, we calculated the global line-of-sight velocity dispersion of isolated and non-isolated stellar systems in MOND for circular orbits. For details on the simulation see HBK09. In Figure~\ref{figtheory}, we plot the two curves from these calculations showing the global line-of-sight velocity dispersion as a function of stellar mass for the classical (open squares) and the modified Newtonian case (open circles). For a given total mass the velocity dispersion in the \mbox{MONDian} case is larger than in the classical theory. In our case, we observed a line-of-sight velocity dispersion (shown as the horizontal lines) and derived the cluster's mass (shown as the vertical lines).

We measured a line-of-sight velocity dispersion of ($0.38~\pm~0.12$)~\mbox{km s$^{-1}$}, not including Star~15. For such a low dispersion, the theoretically predicted mass in MOND is \mbox{$950^{+600}_{-400}~M_\odot$}, and in classical dynamics \mbox{$8\,200^{+6000}_{-4000}~M_\odot$}. We have observed a lower limit of \mbox{(2\,200 $\pm$ 90)~$M_\odot$} (marked in Figure~\ref{figtheory} by the vertical line labeled \emph{observed}) considering only the area within the half-light radius of Pal~14. Already the lower limit excludes the \mbox{MONDian} case, as we have observed more stellar mass than MOND predicts and the stars outside the cluster's half-light radius are not considered yet. The total mass of $\sim$12\,500~$M_\odot$ (in Figure~\ref{figtheory} vertical line marked \emph{extrapolated ($\alpha = 1.3$)}) is several times larger than the \mbox{MONDian} prediction. A declining extrapolation at the low-mass end down to 0.1~$M_\odot$ gives a total mass of Pal~14 of 5\,860~$M_\odot$ (in Figure~\ref{figtheory} vertical line marked \emph{extrapolated ($\alpha = -1$)}), which is also clearly higher than the \mbox{MONDian} prediction. The resulting dynamical mass-to-light ratio for the classical Newtonian case is $M/L_{dyn}$~=~($1.48^{+1.00}_{-0.70}$)~$M_\odot/L_\odot$.

If we include the measured velocity of Star~15, we find a line-of-sight velocity dispersion of ($0.64~\pm~0.15$)~\mbox{km s$^{-1}$}. According to the theoretical calculation of HBK09 the cluster mass in MOND would be $2\,600^{+1400}_{-1200}~M_\odot$, and in classical dynamics $24\,000^{+11000}_{-10000}~M_\odot$. In this case, the extrapolated mass is still larger than the predicted mass in MOND. Also, for the declining mass function for masses $<0.5$~$M_\odot$ the total mass is larger than the \mbox{MONDian} prediction.

Although the measured low velocity dispersion is an indication of whether MOND or classical Newtonian dynamics is correct, one can think of a scenario in which the cluster would be governed by MOND but shows at the same time a velocity dispersion consistent with the classically derived (low) value. n MOND the gravitational force is effectively stronger than in classical dynamics. The stars in a GC which resides in the MOND regime therefore acquire a higher internal velocity, thus leading to a shorter dynamical time and a faster relaxation time for the cluster \citep{ciotti04,zhao05}. Therefore, already after only a couple of orbits around the Galaxy, Pal~14 would have lost a large fraction of its low-mass stars and stellar remnants, leaving the cluster enriched in stars around the main-sequence turnoff and on the red giant branch. In one of Pal~14's perigalactica (if it is on an eccentric orbit), the cluster would become partially unbound and would expand, while it still resides in the classical Newtonian environment close to the Galaxy. The unbound cluster, then, would move further outward on its orbit and would eventually drift into the \mbox{MONDian} regime in the Galaxy's outskirts. As MOND is `stickier' than classical Newtonian dynamics, the stars are bound more strongly again. As a consequence, an observer may measure a low velocity dispersion, similar to the value derived in classical dynamics. At the same time, the cluster's mass is small. For such a scenario to be valid, the cluster would have to be strongly effected by tidal forces, but should not move too far in to be completely destroyed. Detailed simulations on the influence of radial orbits on the velocity dispersion in MOND are necessary.  Unfortunately no proper motion is available for Pal~14 in order to make any constraints on its orbit.

MOND is not the only modification of classical Newtonian dynamics. One other possible theory is modified gravity \citep[MOG;][]{moffat05,moffat08}. MOG explains/predicts galaxy rotation curves, galaxy cluster masses, etc. and at the same time produces predictions consistent with classical dynamics for smaller systems, e.g. GCs. MOG predicts little or no observable deviation from classical Newtonian gravity for GCs with masses of a few times $10^6~M_\odot$ \citep{moffat08}. Our result is consistent with the classical prediction and can, therefore, neither support nor contradict MOG.

\subsection{Velocity dispersion profile and dark matter}

\begin{figure}
\epsscale{1.0}
\plotone{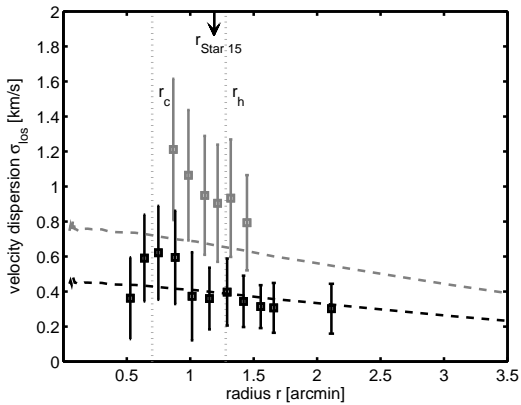}
\caption{The velocity dispersion profile of Pal~14 using running bins with six stars in each bin. The black squares denote the velocity dispersion without Star~15. The gray squares denote these bins where Star~15 was included. The black and gray dashed curves are the theoretical dispersion profiles if Star~15 was included and excluded, respectively. The vertical, dotted lines are the core and half-light radii (H06), respectively. The arrow at the top of the plot marks the radial distance of Star~15 from the cluster center.\label{figdisp}}
\end{figure}

It is widely believed that globular clusters contain no dark matter \citep[e.g.,][]{moore96}. Their dynamical masses closely match the values from population synthesis \citep{mclaughlin05}. The velocity dispersion profile of GCs should, therefore, show a Keplerian fall-off. \citet{scarpa07} studied velocity dispersion profiles of six GCs in the Galaxy. For the five high-concentration clusters in their study ($\omega Cen$, NGC~6171, NGC~6341, NGC~7078, NGC~7099) they found the predicted fall-off in the inner parts of the clusters, but also an unexpected flattening in the outer parts. On the other hand, for the low-concentration cluster NGC~288 they found a more or less flat dispersion profile. For the high-concentration clusters, the profiles always flatten at a radius where the acceleration is around the MONDian limit of $a_0$ for a mass-to-light ratio of 1. To draw any conclusion about MOND from this is rather difficult, as the discussed clusters' total accelerations are not below $a_0$ and therefore the effect of MOND is tiny or even not existent; the clusters are all too close to the Galactic center.

All our stars but one are located within 2.5\arcmin~of the center of Pal~14. We derived the line-of-sight velocity dispersion profile with running radial bins, each bin containing six stars. Figure~\ref{figdisp} shows the resulting velocity dispersion profile. Between 1\arcmin~and 1.5\arcmin~we derived the velocity dispersion either including Star~15 or excluding Star~15. The lower black squares are the case where Star~15 was not included, the upper gray squares the case including Star~15's velocity. For the case excluding Star~15, we can see (within the errors) a slightly declining velocity dispersion profile. The dashed curves in Figure~\ref{figdisp} are the theoretically calculated profiles of HBK09. If we compare our dispersion profile to the theoretical predictions we see a slow fall-off towards outer radii for both. We have observed velocity measurements in the inner 2.5\arcmin~($\sim$50~pc~$\sim$~3.6 core radii). On the other hand, \citet{scarpa07} showed the velocity dispersion profile of NGC~288, another sparse GC with a concentration of c~=~0.96 \citep{harris96}. They describe the profile to be flat out to 4.5 core radii. In order to improve the significance of the comparison of the theoretical prediction and the observational data for Pal~14 as well as of the comparison with similar clusters, spectroscopic data out to larger radii are needed for Pal~14.

We treat this GC the same way as dwarf spheroidal (dSph) galaxies in \citet{madau08} to calculate the central density, using \mbox{$\rho_0 = 166\eta\sigma^2/r_c^2$ $M_\odot pc^{-3}$}, setting $\eta=1$, $r_c=0.7\arcmin =14.5~pc$, and $\sigma=0.38\pm0.12$~\mbox{km s$^{-1}$}. We find a central density of $\rho_{0}=0.1\pm0.07$~$M_\odot pc^{-3}$. A value which is very similar to values found for dSph galaxies (see e.g. Table~1 in \citet{madau08}). On the other hand, if we derive the density within the half-light radius from our mass estimate \mbox{$\mathcal{M}_{tot,hl}$ = (6\,020 $\pm$ 500)~$M_\odot$} and $r_h=1.28\arcmin~=26~pc$, we find $\rho=0.08\pm0.01$~\mbox{$M_\odot pc^{-3}$}. Within the errors the two values agree. We do not need to assume DM for Pal~14.

\section{Summary}

Modified Newtonian dynamics has proven to be quite successful on galactic and also on intergalactic scales \citep{sanders02}. However, not only galaxy size objects must be correctly explained by MOND. Objects with similar low accelerations, for which there is no need for additional, unseen matter such as  GCs must be described correctly by this modified theory, as well. Hence, we have studied the outer halo GC Pal~14 to test whether modified or classical Newtonian dynamics applies. Pal~14 has an internal and external acceleration that are both significantly smaller than $a_0$. Also, the total acceleration of stars in Pal~14 is still significantly smaller than $a_0$ and therefore, Pal~14 is an excellent test object for the two theories. 

We determined the radial velocities of 17 giant stars in Pal~14. Using the measurements of all 17 giants, we confirmed the cluster's mean radial velocity of ($72.19~\pm~0.18$)~\mbox{km s$^{-1}$} and measured a global line-of-sight velocity dispersion of ($0.64~\pm~0.15$)~\mbox{km s$^{-1}$} (see Section~\ref{resultsspec} for details). Excluding Star~15, we find a similar systemic velocity of Pal~14 of ($72.28~\pm~0.12$)~\mbox{km s$^{-1}$} and a lower velocity dispersion of ($0.38~\pm~0.12$)~\mbox{km s$^{-1}$}. These velocity dispersions lead to dynamical masses of \mbox{$950^{+600}_{-400}~M_\odot$} in modified dynamics, and \mbox{$8\,200^{+6000}_{-4000}~M_\odot$} in classical dynamics for the case without Star~15. In the case including Star~15 we expect total masses of Pal~14 of $2\,600^{+1400}_{-1200}~M_\odot$ in MOND, and $24\,000^{+11000}_{-10000}~M_\odot$ in classical dynamics.

The mass function of Pal~14 has a slope of \mbox{$\alpha=1.27~\pm~0.44$} in the mass range $0.53M_\odot$ to $0.78M_\odot$ and is thus flatter than the canonical mass function. This is consistent with the cluster being formed mass segregated with a normal (canonical) IMF but suffering major mass loss through gas expulsion \citep{marks08}. The HST image covers only $7\%$ of the area within the cluster's tidal radius, but more than $2/3$ of the area within the half-light radius. The observed total mass within the half-light radius with an extrapolation to lower masses with a Kroupa-like mass function is $\sim6\,020~M_\odot$. If we extrapolate with a linearly declinig slope for masses $<0.5M_\odot$, we get a total mass within the half-light radius of $\sim2\,930~M_\odot$. In both cases, these values are lower limits. By doubling the numbers to get a rough estimate of the total mass of Pal~14, we get numbers that are substantially higher than the predictions made by HBK09 for MOND. Hence, the cluster's current stellar content is an indication against \mbox{MONDian} dynamics, unless the cluster is on an eccentric orbit.

If Pal~14 is on a circular orbit, MOND cannot explain the low velocity dispersion and the measured mass simultaneously. If Pal~14 is on an eccentric orbit, the low velocity dispersion may still be a problem for MOND, but the measured mass function slope, being flatter than the canonical value, does not allow us to draw a definite conclusion. With the sample of BGK05 and the theoretical predictions of BGK05 and HBK09 we have a basis for extending the study to other outer halo, low-mass Galactic GCs to further refine and improve the tests of gravitational theory. 

\acknowledgments
We like to thank Marina Rejkuba for her advice deriving the radial velocities, and Katharina Glatt for her help with the isochrone fitting and the age determination of Pal~14. KJ and EKG acknowledge support from the Swiss National Foundation through grant numbers 20020-122140 and 20020-113697. SGD acknowledges a partial support from the NSF grant AST-0407448


\end{document}